\documentclass[11pt,a4paper]{article}

\usepackage{amsmath}
\usepackage{amssymb}
\usepackage{amsthm}
\usepackage{multirow}
\usepackage{cite}
\usepackage{listings}
\usepackage{tikz}
\usepackage{pgfplots}
\usepackage{subfigure}
\usepackage{wrapfig}
\usepackage{hyperref}

\newtheorem{thm}{Theorem}
\newtheorem{alg}{Algorithm}
\newtheorem{lemma}{Lemma}

\newcommand{\etal}{\textit{et al.}}
\newcommand{\ie}{\textit{i.e.},}
\newcommand{\eg}{\textit{e.g.},}

\newcommand{\THIMBLEDOCURL}{http://www.stochastik.math.uni-goettingen.de/biometrics/fileadmin/thimble/doc-2014.09.07/}

\newcommand{\BEGINPROOF}{\begin{proof}}
\newcommand{\ENDPROOF}{\end{proof}}

\newcommand{\NumRows}{\text{$n$}}
\newcommand{\NumCols}{\text{$k$}}
\newcommand{\SomeLength}{\text{$\ell$}}

\newcommand{\TheField}{\text{\it\em\bf F}}
\newcommand{\BinaryField}{\text{$\{0,1\}$}}
\newcommand{\bplus}{\text{$~\oplus~$}}

\newcommand{\TheCode}{\text{\it\em\bf C}}

\newcommand{\TheBlockLength}{\text{$n$}}
\newcommand{\TheDimension}{\text{$k$}}
\newcommand{\TheMinimalDistance}{\text{$d$}}
\newcommand{\HammingWeightBound}{\text{$b$}}

\newcommand{\GenMatrix}{\textsl{G}}
\newcommand{\CheckMatrix}{\textsl{H}}
\newcommand{\PermMatrix}{\textsl{P}}

\newcommand{\SomeVector}{\text{$v$}}
\newcommand{\CodeWord}{\text{$c$}}
\newcommand{\Witness}{\text{$w$}}
\newcommand{\Message}{\text{$m$}}
\newcommand{\ErrorPattern}{\text{$e$}}
\newcommand{\NumErrorPatterns}{\text{$B$}}
\newcommand{\FuzzyCommitment}{\text{${\bf f}$}}
\newcommand{\ReceivedWord}{\text{$r$}}

\newcommand{\hash}{\text{$h$}}

\newcommand{\FieldPermutation}{\text{$\sigma$}}

\newcommand{\WitnessEntry}{\text{$u$}}

\newcommand{\UnityVector}{\text{$e$}}
\newcommand{\ZeroVector}{\text{$0$}}
\newcommand{\Transform}{\text{$T$}}
\newcommand{\FirstMatrix}{\textsl{Q}}
\newcommand{\SecondMatrix}{\textsl{R}}

\newcommand{\AlgInput }{\text{ }\newline\textit{\em\bf Input}~~~~}
\newcommand{\AlgOutput}{\text{ }\newline\textit{\em\bf Output}~~~} 
\newcommand{\AlgFailure}{\textsc{Failure}}

\begin{document}

\newcommand{\BenjaminTams}{Benjamin Tams\thanks{B. Tams is with the Institute for Mathematical Stochastics, University of Goettingen, Goldschmidtstr. 7, D-37077, Goettingen, Germany. Phone: +49-(0)551-3913515. Email: \href{mailto:btams@math.uni-goettingen.de}{btams@math.uni-goettingen.de}.}}

\title{Decodability Attack against the\\Fuzzy Commitment Scheme\\with Public Feature Transforms}

\author{\BenjaminTams}

\maketitle

\begin{abstract}
\boldmath
The \emph{fuzzy commitment scheme} is a cryptographic primitive that can be used to store biometric templates being encoded as fixed-length feature vectors protected. If multiple \emph{related} records generated from the same biometric instance can be intercepted, their correspondence can be determined using the \emph{decodability attack}. In 2011, Kelkboom \etal{} proposed to pass the feature vectors through a record-specific but public permutation process in order to prevent this attack. In this paper, it is shown that this countermeasure enables another attack also analyzed by Simoens \etal{} in 2009 which can even ease an adversary to fully break two related records. The attack may only be feasible if the protected feature vectors have a reasonably small Hamming distance; yet, implementations and security analyses must account for this risk. This paper furthermore discusses that by means of a public transformation, the attack cannot be prevented in a binary fuzzy commitment scheme based on linear codes. Fortunately, such transformations can be generated for the non-binary case. In order to still be able to protect binary feature vectors, one may consider to use the \emph{improved fuzzy vault scheme} by Dodis \etal{} which may be secured against linkability attacks using observations made by Merkle and Tams.
\end{abstract}

\section*{Keywords} 
fuzzy commitment scheme, cross-matching, countermeasures

\section{Introduction}

In 1999, Juels and Wattenberg \cite{bib:JuelsWattenberg1999} proposed the \emph{fuzzy commitment scheme} which is considered as an alternative to cryptographic password hashes in which passwords are replaced by samples measured from an individual's biometric characteristic such as, for example, his eyes' \emph{irises} \cite{bib:HaoAndersonDaugman2006}. Given a biometric template the fuzzy commitment scheme can be used to generate a protected record from which it is hard to derive the template unless a sufficiently similar template is presented. Making certain assumptions on the distribution of the biometric templates, the fuzzy commitment scheme provably provides a certain resistance against \emph{irreversibility attacks}. There are, however, more risks than mere irreversibility attack scenarios. For example, the ISO/IEC 24745:2011 international standard \cite{bib:ISO24745:2011} explicitly prohibits that the relation between records generated from the same biometric characteristic can be recognized, \ie{} the \emph{unlinability requirement}. In fact, as for example investigated by Simoens \etal{} \cite{bib:SimoensEtAl2009}, an intruder can link two (or more) related records generated by a fuzzy commitment scheme via the \emph{decodability attack}. In order to prevent the decodability attack, Kelkboom \etal{} \cite{bib:KelkboomEtAl2011} proposed to pass the biometric templates through a record-specific but public permutation process.

On the other hand, the countermeasure makes the scheme vulnerable to a \emph{generalized decodability attack} also considered in \cite{bib:SimoensEtAl2009}. More precisely, even though the performance of decodability attack-based cross-matching is effectively reduced, an adversary may still be able to recognize records protecting very similar templates and may now even recover the biometric templates explicitly. To the best of the author's knowledge, this observation is novel.

\subsection{Contribution and Outline}
After having briefly discussed work that is related to the topic of this paper, in Section \ref{sec:preliminaries} we review the functioning of the fuzzy commitment scheme, repeat the decodability attack, the countermeasure proposed by Kelkboom \etal{} \cite{bib:KelkboomEtAl2011}, and the generalized decodability attack. In Section \ref{sec:theattack} we show how the countermeasure can be exploited by an intruder to run the generalized decodability attack and demonstrate its applicability by reporting experimental results. Furthermore, we discuss problems with designing effective and feasible countermeasures in a binary fuzzy commitment scheme. In Section \ref{sec:countermeasure} we propose a modification of the countermeasure proposed in \cite{bib:KelkboomEtAl2011} that effectively prevents a general class of decodability attacks against non-binary fuzzy commitment schemes. For the binary case, we briefly discuss some simple approaches to fix the problem and finally propose to consider the possibility of using the \emph{improved fuzzy vault scheme} by Dodis \etal{} \cite{bib:DodisEtAl2008} for template protection which, in combination with ideas given by Merkle and Tams \cite{bib:MerkleTams2013}, can be secured against known linkability attacks. Conclusions and an outlook are given in Section \ref{sec:discussion}. The appendix contains an example of reproducing our experiments using our C++ software library that has been made public for download.

\subsection{Related Work}
There are other schemes, for example the \emph{fuzzy vault scheme} \cite{bib:JuelsSudan2002,bib:JuelsSudan2006} that, as the fuzzy commitment scheme, belongs to a general class of protection schemes called \emph{fuzzy sketches} \cite{bib:DodisEtAl2008}. As for the fuzzy commitment scheme, there do also exist attacks via record multiplicity against these schemes \cite{bib:ScheirerBoult2007,bib:KholmatovYanikoglu2008,bib:MerkleTams2013}. Blanton and Aliasgari worked out the fact that none of these can be safely reused unless being encrypted with secret keys \cite{bib:BlantonAliasgari2013}. In fact, using information-theoretic arguments one may argue that, by means of public transformations, the problem of record multiplicity attacks cannot be solved. On the other hand, we stress that these measures still have the potential to render the computational complexity of such attacks infeasible.
\section{Preliminaries} \label{sec:preliminaries}

\subsection{Notation}
Throughout, by $\TheField$ we denote a fixed finite field. By $\TheField^\NumRows$ we denote the set of all column vectors over $\TheField$ of length $\NumRows$; for $\SomeVector\in\TheField^\NumRows$ we denote by $\SomeVector^\top$ the corresponding transposed row vector and by $|\SomeVector|$ we denote the \emph{Hamming weight} of $\SomeVector$, \ie{} the number of positions in $\SomeVector$ that are non-zero. By $\TheField^{\NumRows\times\NumCols}$ we denote the set of all matrices over $\TheField$ with $\NumRows$ rows and $\NumCols$ columns. Given a linear error-correcting code $\TheCode\subset\TheField^\TheBlockLength$, we denote by $\GenMatrix\in\TheField^{\TheBlockLength\times\TheDimension}$ one of its generator matrices such that $\TheCode=\{\GenMatrix\cdot\Message~|~\Message\in\TheField^{\TheDimension}\}$ (note that in the literature there exists also the convention to define the transposed  $\GenMatrix^\top$ as the generator matrix); consequently, we denote a check matrix of $\TheCode$ by a matrix $\CheckMatrix\in\TheField^{(\TheBlockLength-\TheDimension)\times\TheBlockLength}$ of full rank  such that $\CheckMatrix\cdot\GenMatrix=0$.

\subsection{The Fuzzy Commitment Scheme}
In this paper, we consider fuzzy commitment schemes based on linear error-correcting codes over an arbitrary finite field $\TheField$. Let $\TheCode\subset\TheField^\TheBlockLength$ be an $(\TheBlockLength,\TheDimension,\TheMinimalDistance)$-error-correcting code of block length $\TheBlockLength$, dimension $\TheDimension$ and minimal distance $\TheMinimalDistance$ such that it is capable of rounding every $\ReceivedWord\in\TheField^\TheBlockLength$ to its closest codeword $\CodeWord$ provided that $|\CodeWord-\ReceivedWord|\leq (\TheMinimalDistance-1)/2$. For more details on the concept of error-correcting codes we refer to one among the good textbooks available in the literature, for example \cite{bib:Berlekamp1984}. 

Let $\Witness\in\TheField^\TheBlockLength$ be a \emph{feature vector} encoding a biometric template extracted from some biometric modality. A codeword $\CodeWord\in\TheCode$ is selected randomly and is then used to protect $\Witness$ by publishing the \emph{fuzzy commitment} $\FuzzyCommitment=\CodeWord+\Witness$. In order to allow safe recovery of the correct template on genuine verification, we sometimes store a cryptographic hash value $\hash(\CodeWord)$ along with the fuzzy commitment such that $(\FuzzyCommitment,\hash(\CodeWord))$ is considered as the protected record. 

If we assume, for simplicity, that the biometric feature vectors are distributed uniformly among all $\Witness\in\TheField^\TheBlockLength$. Then finding the correct template $\Witness$ from the record $\FuzzyCommitment$ is as hard as iterating through all codewords in $\TheCode$ or, if accessible, reverting the cryptographic hash $\hash(\CodeWord)$ \cite{bib:JuelsWattenberg1999}. It is important to note, that in an analysis the assumption that feature vectors are distributed uniformly among all $\TheField^\TheBlockLength$ typically leads to severe overestimation of system security.

On verification, an allegedly genuine user provides a template $\Witness'\in\TheField^\TheBlockLength$ that is used to be verified against the record $(\CodeWord+\Witness,\hash(\CodeWord))$. If $|\Witness-\Witness'|\leq (\TheMinimalDistance-1)/2$, then $\CodeWord+\Witness-\Witness'$ differs from $\CodeWord$ in not more than $(\TheMinimalDistance-1)/2$ positions, thus can be rounded to $\CodeWord$ using the decoder; the correctness of the codeword $\CodeWord$ can be verified using $\hash(\CodeWord)$ if part of the record, which is deemed to be an accept. Otherwise, if $|\Witness-\Witness'|>(\TheMinimalDistance-1)/2$, the decoder might output a codeword $\CodeWord'$ that (most likely) is different from the correct codeword $\CodeWord$, or fail to decode; both cases are considered as a reject.

\subsection{Decodability Attack} \label{sec:DecodabilityAttack}
Assume that an intruder has intercepted two fuzzy commitments $\FuzzyCommitment_1=\CodeWord_1+\Witness_1$ and $\FuzzyCommitment_2=\CodeWord_2+\Witness_2$ and wants to decide whether they are \emph{related}, \ie{} whether they protect templates $\Witness_1$ and $\Witness_2$ with $|\Witness_1-\Witness_2|\leq (\TheMinimalDistance-1)/2$; note that, the adversary does not necessarily aim at recovering $\Witness_1$ or $\Witness_2$. Therefore, he computes the offset $\ReceivedWord=\FuzzyCommitment_1-\FuzzyCommitment_2=(\CodeWord_1-\CodeWord_2)+(\Witness_1-\Witness_2)$. Since $\TheCode$ is linear, $\CodeWord_1-\CodeWord_2\in\TheCode$. Thus, if $|\Witness_1-\Witness_2|\leq (\TheMinimalDistance-1)/2$, the offset $\ReceivedWord$ can be successfully decoded to a valid codeword; otherwise, if $|\Witness_1-\Witness_2|>(\TheMinimalDistance-1)/2$, the decoding attempt is supposed to output a valid codeword with the non-zero probability $|\TheField|^{\TheDimension-\TheBlockLength}\sum_{j=0}^{\lfloor (\TheMinimalDistance-1)/2\rfloor}(|\TheField|-1)^j{\TheBlockLength\choose j}$ called \emph{sphere packing density} of $\TheCode$. Consequently, if the sphere packing density is small (or even negligible), the adversary can assume that decoding $\ReceivedWord$ yields a valid codeword only if $\FuzzyCommitment_1$ and $\FuzzyCommitment_2$ are related.

Since the sphere packing density is non-zero, the decodability attack might falsely label two non-related records as related. Therefore, the performance of decodability attack-based cross-matching performance is worse than the system's operational performance \cite{bib:SimoensEtAl2009,bib:KelkboomEtAl2011}. On the other hand, if the sphere packing density is negligible, the cross-matching performance can be very close to the operational performance.

\subsection{Prevention via Record-Specific Permutation Processes} \label{sec:PermutationProcess}
In 2011, Kelkboom \etal{} \cite{bib:KelkboomEtAl2011} proposed the incorporation of a public record-specific permutation process to prevent the decodability attack. Therefore, let $\PermMatrix\in\TheField^{\TheBlockLength\times\TheBlockLength}$ be a permutation matrix. Instead of letting $\CodeWord+\Witness$ be the fuzzy commitment, we set $\FuzzyCommitment=\CodeWord+\PermMatrix\Witness$ and publish $(\FuzzyCommitment,\PermMatrix)$ as the record.

Now assume that an adversary has intercepted two records $(\FuzzyCommitment_1,\PermMatrix_1)$ and $(\FuzzyCommitment_2,\PermMatrix_2)$ where $\FuzzyCommitment_1=\CodeWord_1+\PermMatrix_1\Witness_1$ and $\FuzzyCommitment_2=\CodeWord_2+\PermMatrix_2\Witness_2$. Then, since $\PermMatrix_1$ and $\PermMatrix_2$ are random, we can assume that the vectors $\PermMatrix_1\Witness_1$ and $\PermMatrix_2\Witness_2$ are random. For the offset $\ReceivedWord=\FuzzyCommitment_1-\FuzzyCommitment_2$ to be decodable, $|\PermMatrix_1\Witness_1-\PermMatrix_2\Witness_2|\leq (\TheMinimalDistance-1)/2$ must be fulfilled the probability of which can be estimated with the sphere packing density of $\TheCode$. Thus, the cross-matching performance based on direct application of the decodability attack becomes close to random. 

Kelkboom \etal{} \cite{bib:KelkboomEtAl2011} have not considered the following generalization of the decodability attack which, in fact, can be applied by an adversary to determine whether two records protect feature vectors of small difference and even recover them explicitly.

\subsection{Generalized Decodability Attack} \label{sec:GeneralizedDecodabilityAttack}
Simoens \etal{} \cite{bib:SimoensEtAl2009} also considered the case in which the adversary has intercepted two records $\FuzzyCommitment_1=\CodeWord_1+\Witness_1$ and $\FuzzyCommitment_2=\CodeWord_2+\Witness_2$ where $\CodeWord_1$ and $\CodeWord_2$ are codewords from different codes $\TheCode_1$ and $\TheCode_2$, respectively. As before let $\ReceivedWord=\FuzzyCommitment_1-\FuzzyCommitment_2$. Denote by $\GenMatrix_1\in\TheField^{\TheBlockLength\times\TheDimension_1}$ and $\GenMatrix_2\in\TheField^{\TheBlockLength\times\TheDimension_2}$ generator matrices for $\TheCode_1$ and $\TheCode_2$, respectively, let $\tilde{\GenMatrix}=(\GenMatrix_1|\GenMatrix_2)\in\TheField^{\TheBlockLength\times(\TheDimension_1+\TheDimension_2)}$ be the concatenation of the columns of $\GenMatrix_1$ and $\GenMatrix_2$, and write $\ErrorPattern=\Witness_1-\Witness_2$ for the error pattern between the feature vectors. If we write $\tilde{\TheCode}=\{~\tilde{\GenMatrix}\Message~|~\Message\in\TheField^{\TheDimension_1+\TheDimension_2}~\}$ for the code being generated by $\tilde{\GenMatrix}$, then attempting to decode $\ReceivedWord$ in $\tilde{\TheCode}$ leads to a \emph{generalized decodability attack}.

In general it is not clear how to efficiently decode the offset $r$, even if efficient decoders are available for $\TheCode_1$ and $\TheCode_2$. However, if the difference between the feature vectors is small, \ie{} if the error pattern $\ErrorPattern=\Witness_1-\Witness_2$ is of small Hamming weight, then \emph{maximum likelihood decoding} may be a feasible solution to implement the decodability attack. In the following, we describe the procedure in more detail.

Iterate through all $\ErrorPattern_1,\ErrorPattern_2,...\in\TheField^{\TheDimension}$ of Hamming weight less than or equals a small bound $\HammingWeightBound$. If in the $j$th iteration $\ErrorPattern_j=\Witness_1-\Witness_2$, then $\tilde{\CodeWord}=\ReceivedWord-\ErrorPattern_j\in\tilde{\TheCode}$ and we may label $\FuzzyCommitment_1$ and $\FuzzyCommitment_2$ as related; otherwise, if for all tested $\ErrorPattern_j$ we have $\ReceivedWord-\ErrorPattern_j\notin\tilde{\TheCode}$, the records are labeled as non-related.

\subsubsection*{Irreversibility}
If the correct error pattern $\ErrorPattern_j=\Witness_1-\Witness_2$ can be recovered and if the matrix $\tilde{\GenMatrix}$ has rank $\tilde{\TheDimension}=\TheDimension_1+\TheDimension_2$, then we can even fully recover the feature vectors $\Witness_1$ and $\Witness_2$. More specifically, we can find a unique $\tilde{\Message}\in\TheField^{\TheDimension_1+\TheDimension_2}$ solving $\tilde{\CodeWord}=\tilde{\GenMatrix}\tilde{\Message}$. Write $\tilde{\Message}^\top=(\Message_1^\top|-\Message_2^\top)$ where $\Message_1\in\TheField^{\TheDimension_1}$ and $\Message_2\in\TheField^{\TheDimension_2}$. Then, $\CodeWord_1=\GenMatrix_1\Message_1$ and $\CodeWord_2=\GenMatrix_2\Message_2$ (note that, if hash values $\hash(\CodeWord_1)$ and $\hash(\CodeWord_2)$ are stored along with $\FuzzyCommitment_1$ and $\FuzzyCommitment_2$, respectively, the correctness of $\CodeWord_1$ and $\CodeWord_2$ can even be verified). Finally, the feature vectors can be determined via the relations $\Witness_1=\FuzzyCommitment_1-\CodeWord_1$ and $\Witness_2=\FuzzyCommitment_2-\CodeWord_2$.

If $\tilde{\GenMatrix}$ has rank $\tilde{\TheDimension}<\TheDimension_1+\TheDimension_2$, the solution of $\tilde{\Message}$ is not unique. However, there exists such an $\tilde{\Message}$ yielding the correct $\CodeWord_1$ and $\CodeWord_2$. Therefore, depending on the rank of $\tilde{\GenMatrix}$, the adversary gains advantage for fully recovering feature vectors $\Witness_1$ and $\Witness_2$ of small Hamming distance from $\FuzzyCommitment_1$ and $\FuzzyCommitment_2$. 

\section{Decodability Attack in a Fuzzy Commitment Scheme with Public Record-Specific Permutation} \label{sec:theattack}

In this section, we consider a simple transformation from the problem of cross-matching two records generated by the fuzzy commitment scheme of which feature vectors have been passed through public record-specific permutation processes to the problem of cross-matching two records generated by the fuzzy commitment scheme based on different codes.

\subsection{The Attack}
Let $(\FuzzyCommitment_1,\PermMatrix_1)$ and $(\FuzzyCommitment_2,\PermMatrix_2)$ be two records generated with a fuzzy commitment scheme based on an $(\TheBlockLength,\TheDimension,\TheMinimalDistance)$-code $\TheCode$ over the field $\TheField$. More precisely, $\FuzzyCommitment_1=\CodeWord_1+\PermMatrix_1\Witness_1$, $\FuzzyCommitment_2=\CodeWord_2+\PermMatrix_2\Witness_2$, where $\CodeWord_1,\CodeWord_2\in\TheCode$ and $\Witness_1,\Witness_2\in\TheField^\TheBlockLength$ both encode biometric templates.

Let $\GenMatrix\in\TheField^{\TheBlockLength\times\TheDimension}$ be a generator matrix of the code $\TheCode$ and set $\GenMatrix_1=\PermMatrix_1^{-1}\GenMatrix$ and $\GenMatrix_2=\PermMatrix_2^{-1}\GenMatrix$. Denote by $\TheCode_1$ and $\TheCode_2$ the codes being generated by $\GenMatrix_1$ and $\GenMatrix_2$, respectively. Furthermore, write $\CodeWord'_1=\PermMatrix_1^{-1}\CodeWord_1$ and $\CodeWord'_2=\PermMatrix_2^{-1}\CodeWord_2$ which are codewords in $\TheCode_1$ and $\TheCode_2$, respectively. Then $\PermMatrix_1^{-1}\FuzzyCommitment_1=\CodeWord'_1+\Witness_1$ and $\PermMatrix_2^{-2}\FuzzyCommitment_2=\CodeWord'_2+\Witness_2$ are fuzzy commitments constructed over $\TheCode_1$ and $\TheCode_2$, respectively. This situation is equivalent to the attack scenario of Section \ref{sec:GeneralizedDecodabilityAttack} and leads to the following attack.

\begin{alg}[Modified Decodability Attack]
\AlgInput A linear code $\TheCode\subset\TheField^\TheBlockLength$ given by its generator matrix $\GenMatrix\in\TheField^{\TheBlockLength\times\TheDimension}$; two records $(\FuzzyCommitment_1,\PermMatrix_1)$ and $(\FuzzyCommitment_2,\PermMatrix_2)$ where $\FuzzyCommitment_1$ and $\FuzzyCommitment_2$ are of the form $\CodeWord_1+\PermMatrix_1\Witness_1$ and $\CodeWord_2+\PermMatrix_2\Witness_2$, respectively, where $\CodeWord_1,\CodeWord_2\in\TheCode$, $\Witness_1,\Witness_2\in\TheField^\TheBlockLength$ and permutation matrices $\PermMatrix_1,\PermMatrix_2\in\TheField^{\TheBlockLength\times\TheBlockLength}$; an integer $\HammingWeightBound\geq 0$.
\AlgOutput Either a candidate for $(\Witness_1,\Witness_2)$ or \AlgFailure.
\label{alg:theattack}
\end{alg}
\begin{enumerate}
\item\textbf{[Code offset]} Let $\ReceivedWord=\PermMatrix_1^{-1}\FuzzyCommitment_1-\PermMatrix_2^{-1}\FuzzyCommitment_2$.
\item\textbf{[Check Matrix]} Compute $\GenMatrix_1=\PermMatrix_1^{-1}\GenMatrix$, $\GenMatrix_2=\PermMatrix_2^{-1}\GenMatrix$ and set $\tilde{\GenMatrix}=(\GenMatrix_1|\GenMatrix_2)$; find a matrix of maximal rank such that $\tilde{\CheckMatrix}\cdot\tilde{\GenMatrix}=0$, \eg{} by computing a basis for the kernel of $\tilde{\GenMatrix}^\top$.
\item\textbf{[Find error pattern]} Let $\ErrorPattern_1,...,\ErrorPattern_\NumErrorPatterns\in\TheField^\TheBlockLength$ be all vectors of Hamming weight $\leq\HammingWeightBound$ sorted w.r.t. their Hamming weight, \ie{} $|\ErrorPattern_1|\leq...\leq|\ErrorPattern_\NumErrorPatterns|$. Next, successively iterate through the list until a $j$ is found such that $\tilde{\CheckMatrix}\cdot(\ReceivedWord-\ErrorPattern_j)=0$. If none such $j$ exists, return \AlgFailure{}; otherwise continue with the next step.
\item\label{step:LinearSystem}\textbf{[Solve linear system]} Compute $\tilde{\CodeWord}=\ReceivedWord-\ErrorPattern_j$ and solve $\tilde{\GenMatrix}\tilde{\Message}=\tilde{\CodeWord}$ for $\tilde{\Message}\in\TheField^{\TheDimension+\TheDimension}$. Write $\tilde{\Message}^\top=(\Message_1^\top,-\Message_2^\top)$ where $\Message_1,\Message_2\in\TheField^{\TheDimension}$ and set $\CodeWord_1^*=\PermMatrix_1\GenMatrix\Message_1$ and $\CodeWord_2^*=\PermMatrix_2\GenMatrix\Message_2$.
\item\textbf{[Return]} Determine $\Witness_1^*=\PermMatrix_1^{-1}\cdot(\FuzzyCommitment_1-\CodeWord_1^*)$ and $\Witness_2^*=\PermMatrix_2^{-1}\cdot(\FuzzyCommitment_2-\CodeWord_2^*)$ and return $(\Witness_1^*,\Witness_2^*)$.
\end{enumerate}
For an analysis of the attack, which can be formulated as the attack described in Section \ref{sec:GeneralizedDecodabilityAttack}, we refer to Section 5.1 in \cite{bib:SimoensEtAl2009}.

\subsection{Remarks}
Note that Algorithm \ref{alg:theattack} always returns a feature vector pair if the matrix $\tilde{\GenMatrix}$ has rank $\TheBlockLength$. In this case the outputs may not be very useful. This, however, assumes that $2\TheDimension\geq\TheBlockLength$ which may not be fulfilled for codes that can correct a reasonable number of errors (for example, see the codes listed in Table I in \cite{bib:KelkboomEtAl2011}). Also note that, if a cryptographic hash value is stored along with a fuzzy commitment, the attack can easily be modified to only output the correct feature vectors or \AlgFailure.

\subsection{Experiments} \label{sec:experiments}

\begin{table}[!t]
\begin{center}
\caption{Performance of the attack (Algorithm \ref{alg:theattack}) determined experimentally on a 1.8 GHz server with sufficient memory} \label{tab:experiments}
\begin{tabular}{|c|c||l|c|c|}
\hline
 \multirow{2}{*}{$(\TheBlockLength,\TheDimension,\TheMinimalDistance)$} & \multirow{2}{*}{$\HammingWeightBound$} & rel./non-rel. & rel./non-rel.      & rel./non-rel.         \\
                                                                        &                       & linkage rate   & rec. rate & attack time           \\
\hline\hline
 \multirow{6}{*}{$(31,11,11)$}                                          & $0$                   & 100\% / 0.06\% & 51.2\% / 0\%  & $0ms$ / $0ms$          \\ 
                                                                        & $1$                   & 100\% / 2.94\% & 49.3\% / 0\%  & $0ms$ / $0ms$          \\
                                                                        & $2$                   & 100\% / 39.1\% & 45.7\% / 0\%  & $0ms$ / $0ms$          \\
                                                                        & $3$                   & 100\% / 99.2\% & 25.2\% / 0\%  & $0ms$ / $0ms$          \\
                                                                        & $4$                   & 100\% / ~100\% & 6.36\% / 0\%  & $0ms$ / $0ms$          \\
                                                                        & $5$                   & 100\% / ~100\% & 0.94\% / 0\%  & $0ms$ / $0ms$          \\
\hline\hline
 \multirow{6}{*}{$(63,24,15)$}                                          & $0$                   & 100\% / ~~~0\% & 50.9\% / 0\%  & $0ms$ / $0ms$          \\ 
                                                                        & $1$                   & 100\% / 0.06\% & 49.5\% / 0\%  & $0ms$ / $0ms$          \\
                                                                        & $2$                   & 100\% / 2.82\% & 49.9\% / 0\%  & 1$ms$ / 1$ms$    \\
                                                                        & $3$                   & 100\% / 46.1\% & 43.1\% / 0\%  & 3$ms$ / 8$ms$    \\
                                                                        & $4$                   & 100\% / 99.9\% & 20.3\% / 0\%  & 9$ms$ / 14$ms$   \\
                                                                        & $5$                   & 100\% / ~100\% & 3.26\% / 0\%  & 14$ms$ / 14$ms$  \\
\hline\hline
 \multirow{6}{*}{$(127,50,27)$}                                         & $0$                   & 100\% / ~~~0\% & 50.2\% / 0\%  & 4$ms$ / 4$ms$    \\ 
                                                                        & $1$                   & 100\% / ~~~0\% & 49.8\% / 0\%  & 4$ms$ / 4$ms$    \\
                                                                        & $2$                   & 100\% / ~~~0\% & 50.8\% / 0\%  & 5$ms$ / 7$ms$    \\
                                                                        & $3$                   & 100\% / 0.08\% & 50.0\% / 0\%  & 35$ms$ / 137$ms$ \\
                                                                        & $4$                   & 100\% / 4.18\% & 50.5\% / 0\%  & 932$ms$ / 3.7$s$ \\
                                                                        & $5$                   & 100\% / 62.5\% & 40.9\% / 0\%  & 16$s$ / 54$s$    \\
\hline\hline
 \multirow{5}{*}{$(255,87,53)$}                                         & $0$                   & 100\% / ~~~0\% & 50.3\% / 0\%  & 29$ms$ / 28$ms$  \\ 
                                                                        & $1$                   & 100\% / ~~~0\% & 50.9\% / 0\%  & 29$ms$ / 28$ms$  \\
                                                                        & $2$                   & 100\% / ~~~0\% & 50.2\% / 0\%  & 35$ms$ / 54$ms$  \\
                                                                        & $3$                   & 100\% / ~~~0\% & 50.6\% / 0\%  & 465$ms$ / 1.9$s$ \\
                                                                        & $4$                   & 100\% / ~~~0\% & 49.9\% / 0\%  & 25$s$ / 106$s$   \\
                                                                        & $5$                   & 100\% / ~~~0\% & 48.3\% / 0\%  & 21$min$ / 81$min$   \\
\hline
\end{tabular}
\end{center}
\end{table}

To show that Algorithm~\ref{alg:theattack} can successfully attack two related records generated with a fuzzy commitment scheme we conducted some experiments with an own C++ implementation which has been made public for download; see the appendix (Section \ref{sec:appendix}) for more details. Therein, we considered binary BCH codes \cite{bib:Berlekamp1984} of block length $\TheBlockLength=31,63,127,255$. The codes were designed such that they had a minimal distance being capable of correcting 10\% of bit errors. Furthermore, we considered  Hamming weight bounds $\HammingWeightBound$ as input to the algorithm varying between 0 and 5.

For each tested $(\TheCode,\HammingWeightBound)$ we distinguished two cases: In the first case, we selected feature vector pairs uniformly at random having a Hamming distance less than or equals $\HammingWeightBound$; in the second case, we selected feature vector pairs uniformly at random without accounting for the Hamming distance. To the first case we refer as the \emph{related case} and the second as the \emph{non-related case}. For each feature vector pair, we selected two bit permutations and two codewords uniformly and independently at random. The entries of the feature vectors were reordered using the permutations and then xored with the selected codewords to obtain two fuzzy commitments. The code's generator matrix, the fuzzy commitments together with their respective permutation processes, and the Hamming weight bound $\HammingWeightBound$ were input to our implementation of Algorithm~\ref{alg:theattack}. Whenever the algorithm output candidates for the feature vectors, we accounted the event as a \emph{linkage}. If, in addition, the pair was correct, we accounted the event as a \emph{recovery}. For most tested $(\TheCode,\HammingWeightBound)$ we simulated $5000$ related and $5000$ non-related cases; for $\TheBlockLength=255$ and $\HammingWeightBound=5$ only $350$ related and $350$ non-related cases have been simulated. Thereby, we measured the \emph{average linkage rate}, \emph{average recovery rate} and the \emph{average attack time} for the related and non-related case. The results can be found in Table \ref{tab:experiments}.

Our tests clearly indicate that the fuzzy commitment scheme remains vulnerable to record multiplicity attacks even if the protected feature vectors have been passed through record-specific public permutation processes. For example, for $(\TheBlockLength,\TheDimension)=(127,50)$, according to our tests it is possible to quite reliably distinguish records that protect feature vectors differing in at most $\HammingWeightBound=3$ bit positions from those protecting random feature vectors: In all related cases the attack output a pair, but only in approximately $0.08\%$ of the non-related cases. The time that an adversary has to spent before he can dismiss a non-related record pair was measured as $137ms$. Furthermore, in nearly $50\%$ of the related cases the attack returned the correct feature vectors: This is due to the fact that the linear systems solved in Step \ref{step:LinearSystem} of Algorithm \ref{alg:theattack} had in average two solutions (\ie{} the rank of $\tilde{\GenMatrix}$ was $2\TheDimension-1$) of which only one was correct. In fact, in all cases for $\TheBlockLength=63,127,255$ observed the linear systems had exactly two solutions (in some cases for $\TheBlockLength=31$ the systems had four solutions). This explains why we measured an approximate maximum of $50\%$ related recovery rate. We stress that, if a cryptographic hash value is accessible along with a fuzzy commitment, then the correct of the solution can be selected in Step \ref{step:LinearSystem} and then a related recovery rate of up to $100\%$ can be achieved.

It is important to note the following. Kelkboom \etal{} \cite{bib:KelkboomEtAl2011} considered BCH codes that are able to correct $\approx 25\%$ of errors while we consider codes of $\approx 10\%$ error-correcting capability. We stress that the higher the error-correcting capability of a linear code, the lower is its dimension and thus its sphere packing density; consequently, the attack will even be more effective against the codes considered in \cite{bib:KelkboomEtAl2011}. This is in accordance with the results one achieves using the C++ code example given in the appendix (Section \ref{sec:appendix}).

\subsection{The Binary Case}

Our observations and experiments lead to the question whether it is possible to pass the feature vectors through another class of transformation than processes reordering the position of the feature vectors in order to prevent the (generalized) decodability attack. Such class of transforms should preserve the Hamming distance between two feature vectors in order to not affect the verification performance of the system. More specifically, we say that a map $\Transform:\TheField^\TheBlockLength\rightarrow\TheField^\TheBlockLength$ \emph{preserves the Hamming distance} if for all $\SomeVector_1,\SomeVector_2\in\TheField$ the equality $|\Transform(\SomeVector_1)-\Transform(\SomeVector_2)|=|\SomeVector_1-\SomeVector_2|$ holds. In Section \ref{sec:countermeasure} we construct such transforms and show that they effectively prevent decodability attacks provided $|\TheField|$ is of sufficient size. On the other hand, the use of these transforms may be ineffective for very small finite fields, in particular, if $|\TheField|=2$. The binary case, however, is a very important one and is therefore considered in this section. Unfortunately, the result of our analysis is that \emph{decodability attacks cannot be prevented in the binary case} using a public and record-specific transformation preserving the Hamming distance because every such transformation essentially is a bit permutation process. More precisely:

\begin{thm} \label{thm:BFCS}
Let $\Transform:\BinaryField^\TheBlockLength\rightarrow\BinaryField^\TheBlockLength$ be a map that preserves the Hamming distance. Then there exists a permutation matrix $\PermMatrix\in\BinaryField^{\TheBlockLength\times\TheBlockLength}$ such that $\Transform(\SomeVector)=\PermMatrix\SomeVector\bplus\Transform(\ZeroVector)$ for every $\SomeVector\in\BinaryField^\TheBlockLength$.
\end{thm}

Here we denote by $\BinaryField$ the field with two elements and by $\bplus$ the addition of vectors with coefficients in $\BinaryField$ which can be interpreted as an exclusive or-operation (note that subtraction is the same as addition over $\BinaryField$).

By Theorem \ref{thm:BFCS}, every transformation through which feature vectors can be passed are permutations plus a constant shifting vector. Due to the publicity of the transformations an adversary can subtract the constant shift from any fuzzy commitment that he has intercepted. Then Algorithm \ref{alg:theattack} can be applied to attack related fuzzy commitments as before.

\subsubsection*{Proof of Theorem \ref{thm:BFCS}}
First, note that $\Transform$ must be bijective (as a map preserving the Hamming distance) since, otherwise, there would exist distinct vectors $\SomeVector_1,\SomeVector_2\in\BinaryField^\TheBlockLength$ with $0=|\Transform(v_1)\bplus\Transform(\SomeVector_2)|=|\SomeVector_1\bplus\SomeVector_2|\neq 0$ which is a contradiction.

Let $\hat{\Transform}:\BinaryField^\TheBlockLength\rightarrow\BinaryField^\TheBlockLength,~ \SomeVector\mapsto\Transform(\SomeVector)\bplus\Transform(\ZeroVector)$. Note that $|\SomeVector\bplus\SomeVector'|=|\Transform(\SomeVector)\bplus\Transform(\SomeVector')|=|\Transform(\SomeVector)\bplus\Transform(\SomeVector')\bplus\Transform(\ZeroVector)\bplus\Transform(\ZeroVector)|=|\hat{\Transform}(\SomeVector)\bplus\hat{\Transform}(\SomeVector')|$; consequently, $\hat{\Transform}$ preserves the Hamming distance, too, and is thus bijective as well. Furthermore, since $|\SomeVector|=|\SomeVector\bplus\ZeroVector|=|\Transform(\SomeVector)\bplus\Transform(\ZeroVector)|=|\hat{\Transform}(\SomeVector)|$, $\hat{\Transform}$ also preserves the Hamming weight.

It remains to show that the map $\hat{\Transform}$ is a bit permutation process to which the following lemma is a key.

\begin{lemma} \label{lemma:BitPermutation}
Let $\hat{\Transform}:\BinaryField^\TheBlockLength\rightarrow\BinaryField^\TheBlockLength$ be a map that preserves the Hamming distance and the Hamming weight. For every list of distinct unity vectors $\UnityVector_1,...,\UnityVector_{\SomeLength}\in\BinaryField^\TheBlockLength$, $0\leq\SomeLength\leq\TheBlockLength$, the equality $\hat{\Transform}(\UnityVector_1\bplus\hdots\bplus\UnityVector_\SomeLength)=\hat{\Transform}(\UnityVector_1)\bplus\hdots\bplus\hat{\Transform}(\UnityVector_\SomeLength)$ holds.
\end{lemma}
\BEGINPROOF
From the preservation of the Hamming weight it follows that $\hat{\Transform}(\ZeroVector)=\ZeroVector$, so the statement holds for $\SomeLength=0$. Furthermore, the mappings $\hat{\Transform}(\UnityVector_1),...,\hat{\Transform}(\UnityVector_\SomeLength)$ are unity vectors and pairwise distinct (since $\hat{\Transform}$ preserves the Hamming weight and is bijective). Note that $|\hat{\Transform}(\UnityVector_1\bplus\hdots\bplus\UnityVector_\SomeLength)\bplus\hat{\Transform}(\UnityVector_j)|=\SomeLength-1$ and since the $\hat{\Transform}(\UnityVector_j)$ are all distinct, we obtain
\begin{align*}
\SomeLength&>|\hat{\Transform}(\UnityVector_1\bplus\hdots\bplus\UnityVector_\SomeLength)\bplus\hat{\Transform}(\UnityVector_1)|\\
           &>|\hat{\Transform}(\UnityVector_1\bplus\hdots\bplus\UnityVector_\SomeLength)\bplus\hat{\Transform}(\UnityVector_1)\bplus\hat{\Transform}(\UnityVector_2)|\\
           &~~\vdots\\
           &>|\hat{\Transform}(\UnityVector_1\bplus\hdots\bplus\UnityVector_\SomeLength)\bplus\hat{\Transform}(\UnityVector_1)\bplus\hdots\bplus\hat{\Transform}(\UnityVector_\SomeLength)|=0
\end{align*}
and the statement of the lemma follows.
\ENDPROOF
Due to the fact that every vector in $\BinaryField^\TheBlockLength$ can be written as the sum of unity vectors and applying Lemma \ref{lemma:BitPermutation} recursively, it follows that $\hat{\Transform}$ is linear and since $\hat{\Transform}$ maps unity vectors to unity vectors, it must be a bit permutation process.\hfill$\square$

\section{Countermeasure} \label{sec:countermeasure}
In this section, we show which class of transformations can be used such that decodability attacks can be avoided. The modification may be useful in a fuzzy commitment scheme working with sufficiently large fields ($|\TheField|\geq 23$, say), but is useless for the binary case.

\subsection{Record-Specific Field Permutation Process} \label{sec:FieldPermutation}

Let $\Witness\in\TheField^\TheBlockLength$ encode a template which we want to protect via the fuzzy commitment scheme. As usual, let $\CodeWord\in\TheCode$ be a codeword of a linear $(\TheBlockLength,\TheDimension)$-code $\TheCode\subset\TheField^\TheBlockLength$. Prior to the generation of a fuzzy commitment, instead of passing $\Witness$ through a random process that reorders the positions of the vector $\Witness$, we propose to pass the entries of $\Witness$ through a random bijection $\FieldPermutation:\TheField\rightarrow\TheField$. More precisely, write $\Witness=(\WitnessEntry_1,...,\WitnessEntry_{\TheBlockLength})^\top$ with $\WitnessEntry_j\in\TheField$; then we may choose $\Transform:\TheField^\TheBlockLength\rightarrow\TheField^\TheBlockLength,~(\WitnessEntry_1,...,\WitnessEntry_{\TheBlockLength})^\top\mapsto(\FieldPermutation(\WitnessEntry_1),...,\FieldPermutation(\WitnessEntry_{\TheBlockLength}))^\top$ as the transformation. $\Transform$ obviously preserves the Hamming distance between feature vectors and can be used instead of a process reordering the feature vectors' positions as proposed in \cite{bib:KelkboomEtAl2011}.

\subsection{Linear Decodability Attacks} \label{sec:LinearDecodabilityAttacks}
To analyze the effectiveness of using a record-specific transformation as above, we introduce a class of attacks that contain the (generalized) decodability attack as special cases. Let $\FuzzyCommitment_1=\CodeWord_1+\Transform_1(\Witness_1)$ and $\FuzzyCommitment_2=\CodeWord_2+\Transform_2(\Witness_2)$ be two records generated by the fuzzy commitment scheme where the feature vectors have been passed through public record-specific transformations $\Transform_1$ and $\Transform_2:\TheField^{\TheBlockLength}\rightarrow\TheField^{\TheBlockLength}$, respectively. We define a \emph{linear decodability attack} to be an attack that exploits two invertible $\FirstMatrix,\SecondMatrix\in\TheField^{\TheBlockLength\times\TheBlockLength}$ (that may depend on $\Transform_1$ and $\Transform_2$) such that the Hamming distance between any feature vectors $\Witness_1$ and $\Witness_2$ is preserved as follows: $|\Witness_1-\Witness_2|=|\FirstMatrix\cdot\Transform_1(\Witness_1)-\SecondMatrix\cdot\Transform_2(\Witness_2)|$. Then from decodability of the offset $\FirstMatrix\cdot\FuzzyCommitment_1-\SecondMatrix\cdot\FuzzyCommitment_2$ in the code generated by the matrix $\tilde{\GenMatrix}=(\FirstMatrix\GenMatrix|\SecondMatrix\GenMatrix)$ related $\FuzzyCommitment_1$ and $\FuzzyCommitment_2$ can be recognized and even be broken with the generalized decodability attack described in Section \ref{sec:GeneralizedDecodabilityAttack}. 

For the decodability attack (Section \ref{sec:DecodabilityAttack}) the transforms $\Transform_1$ and $\Transform_2$ and the matrices $\FirstMatrix$ and $\SecondMatrix$ are identities. For the decodability attack against the fuzzy commitment scheme with public record-specific permutation process (Algorithm \ref{alg:theattack}), the transforms $\Transform_1$ and $\Transform_2$ are the linear maps induced by the permutation matrices $\PermMatrix_1$ and $\PermMatrix_2$, respectively, while $\FirstMatrix=\PermMatrix_1^{-1}$ and $\SecondMatrix=\PermMatrix_2^{-1}$.

\subsection{Analysis of the Countermeasure}
We next show that the probability for an effective linear decodability attack to exist can be very small if the entries of the feature vectors have been passed through a public, random and record-specific field permutation. Therefore, note that, in order to ease notation, it is sufficient to assume that only one (the second, say) record protects a feature vector whose entries have been transformed. The key of our estimation is given by the following.
\begin{lemma}
Let $\Transform:\TheField\rightarrow\TheField$ be a map. Assume that there exists invertible matrices $\FirstMatrix,\SecondMatrix\in\TheField^{\TheBlockLength}$ with $|\Witness_1-\Witness_2|=|\FirstMatrix\cdot\Witness_1-\SecondMatrix\cdot\Transform(\Witness_2)|$ for every $\Witness_1,\Witness_2\in\TheField^{\TheBlockLength}$. Then $\Transform$ is linear.
\end{lemma}
\BEGINPROOF
The condition $|\Witness_1-\Witness_2|=|\FirstMatrix\cdot\Witness_1-\SecondMatrix\cdot\Transform(\Witness_2)|$ for every $\Witness_1,\Witness_2\in\TheField^{\TheBlockLength}$ assumes that $0=\FirstMatrix\cdot\Witness-\SecondMatrix\cdot\Transform(\Witness)$ for all $\Witness\in\TheField^{\TheBlockLength}$. Hence $\Transform(\Witness)=\SecondMatrix^{-1}\cdot\FirstMatrix\cdot\Witness$ and the statement of the lemma follows.
\ENDPROOF

For a bijective map $\FieldPermutation:\TheField\rightarrow\TheField$ set $\Transform:\TheField^{\TheBlockLength}\rightarrow\TheField^{\TheBlockLength},~(\WitnessEntry_1,...,\WitnessEntry_{\TheBlockLength})^\top\mapsto(\FieldPermutation(\WitnessEntry_1),...,\FieldPermutation(\WitnessEntry_{\TheBlockLength}))^\top$. Then, if $|\Witness_1-\Witness_2|=|\FirstMatrix\cdot\Witness-\SecondMatrix\cdot\Transform(\Witness_2)|$ it follows from the lemma that $\Transform$ must be linear and thus its components, \ie{} the map $\FieldPermutation:\TheField\rightarrow\TheField$, must be linear, too, and, as a bijection, additionally be invertible.

There are $(|\TheField|-1)\cdot|\TheField|$ invertible linear maps and $|\TheField|!$ permutations. Consequently, we estimate the probability that the incorporation of two random field permutation processes enables a linear decodability attack as $1/(|\TheField|-2)!~$.

For example, for $|\TheField|=32$, $64$ and $128$, these probabilities evaluate to approximately $2^{-108}$, $2^{-284}$ and $2^{-702}$, respectively, which may be reasonably small to prevent an attacker from successfully running a linear decodability attack.

\subsubsection*{Disclaimer}
At this point we stress that, although linear decodability attacks (as considered in this section) can be avoided by incorporating a public field permutation process, there might exist other effective record multiplicity attacks. Yet, to the best of the author's knowledge, there is no known approach being more efficient than breaking the records individually---even in the case of equal feature vectors. However, future research is needed to confirm or disprove the validity of this assumption.

\subsection{The Binary Case} \label{sec:BinaryCountermeasure}
The countermeasure proposed in Section \ref{sec:FieldPermutation} is useful only if we can assume that the fields have a sufficient size. However, many applications work with binary feature vectors. On the other hand, we have discussed that preventing decodability attacks via a public transformation preserving the Hamming distance can be problematic in a binary fuzzy commitment scheme.

One solution to the problem would be to use non-linear error-correcting codes. Then the decodability attack cannot be applied which exploits linearity of the underlying error-correcting code. However, the countermeasure proposed by Kelkboom \etal{} \cite{bib:KelkboomEtAl2011} was meant to address the most dominant linear case. 

In the binary and linear case, in combination with record-specific but public permutation processes, it is possible to flip a few bits randomly such that the error pattern that an adversary needs to correctly guess during the attack is expected to have at least a certain Hamming weight. In such a way the attack might be rendered infeasible. On the other hand, the flipped bits introduce random errors that need to be compensated by a higher error-correction capability of the code and thus a lower dimension which negatively affects irreversibility of the records. Yet, if the dimension is still sufficient to guarantee a certain security, randomly flipping a few bits may be a valid countermeasure for the binary case. Nonetheless, other promising solutions do exist.

\subsubsection*{Switch the Scheme}
Another way to achieve resistance against linkability attacks for binary feature vectors may be to map them to \emph{feature sets}, \ie{} subsets of a finite field, which can be protected using a \emph{fuzzy vault scheme} \cite{bib:JuelsSudan2002}. Therein, each position of the feature vector is attached with a unique field element and for each position in the feature vector that is set to one the feature set is defined to contain the field element attached to the bit position; this relation has already been outlined in \cite{bib:DodisEtAl2008}. The feature set can then be protected using the \emph{improved fuzzy vault scheme} by Dodis \etal{} \cite{bib:DodisEtAl2008}. It is important to note that there exist record multiplicity attacks against the improved fuzzy vault scheme as, too \cite{bib:MerkleTams2013}; in fact, each known biometric template protection scheme is in principle vulnerable to linkability attacks \cite{bib:BlantonAliasgari2013}. However, since the improved fuzzy vault scheme may work with fields containing more than two elements it is possible to pass the feature elements through a record-specific field permutation process \cite{bib:MerkleTams2013} similar to the measure briefly discussed above (Section \ref{sec:FieldPermutation}) in a non-binary fuzzy commitment scheme. It is important to note that when switching to the improved fuzzy vault scheme, fulfilling the unlinkability requirement may not be a the cost of verification performance or security and is therefore, in our view, a very promising fix to the problem. Yet, a proof that passing feature elements through a record-specific (but public) field permutation can effectively prevent any linkability attack would be desirable, \eg{} by proving a complexity-theoretical statement such as ``the publicity of the permutation processes cannot be exploited to gain advantage in linking two records as compared when breaking one the records individually''. Alternatively, we could try to disprove the conjecture by finding an effective linkability attack as a counterexample.
\section{Discussion} \label{sec:discussion}

\subsection{Summary}
We discussed the effect of preventing decodability attack-based cross-matching in a fuzzy commitment scheme by incorporating a random public bit permutation process \cite{bib:KelkboomEtAl2011}. We found that the measure does not completely solve the problem and even can make the scheme vulnerable to reversibility attacks, if the protected feature vectors have a sufficiently small Hamming distance. We supported this observation experimentally and complemented our work by proving that there exists no class of transformations preserving the Hamming distance between feature vectors that solve the problem of decodability attacks in a binary fuzzy commitment scheme. On the other hand, we were able to construct effective transformations provided the underlying finite field is of a sufficient size. For the important binary case, we gave arguments that the improved fuzzy vault scheme by Dodis \etal{} \cite{bib:DodisEtAl2008}, in combination with ideas established by Merkle and Tams \cite{bib:MerkleTams2013}, is a possible alternative to protect binary feature vectors in an unlinkable and key-less biometric template protection system. 

\subsection{Conclusion}
Our work clearly states that the mere incorporation of a record-specific bit permutation process is not sufficient in order to make a binary fuzzy commitment scheme resistant against linkability attacks unless the ineffectiveness of Algorithm~\ref{alg:theattack} can be guaranteed in the security analysis of a specific implementation: For example, by estimating the probability as negligible that related binary biometric feature vectors agree in less than a certain number of bits (depending on the aimed security level) or by guaranteeing that the code $\tilde{\TheCode}$ in the algorithm has a sufficiently large sphere packing density. If such assertions cannot be made, the protection of binary feature vectors with a fuzzy commitment scheme seems not to be possible if based on linear error-correcting codes. Even though we discussed some possible approaches in order to prevent decodability attacks, future research is needed to confirm whether they yield resistance against heavy attacks. 

\subsection{Outlook}
In particular, we briefly discussed the use of the improved fuzzy vault scheme by Dodis \etal{} \cite{bib:DodisEtAl2008} as a possible alternative to implement a key-less and unlinkable protection for binary feature vectors: Even though there exists an effective attack in the presence of record multiplicity against the improved fuzzy vault scheme, there also exists a possible countermeasure \cite{bib:MerkleTams2013} similar to the one discussed in Section \ref{sec:FieldPermutation}. As part of our future research, we plan to prove (or disprove) the infeasibility of attacking the improved fuzzy vault scheme (with countermeasure) from record multiplicity using complexity-theoretical arguments.

\section*{ACKNOWLEDGEMENT}
The support of the Felix Bernstein Institute for Mathematical Statistics in the Biosciences and the Volkswagen Foundation is gratefully acknowledged.

\bibliographystyle{IEEEtran}
\bibliography{IEEEabrv,literature}

\section{Appendix} \label{sec:appendix}

We provide a C++ software \href{\THIMBLEDOCURL/index.html}{THIMBLE} that can be downloaded from
\begin{center}
\footnotesize{\url{http://www.stochastik.math.uni-goettingen.de/biometrics/thimble}}
\end{center}
and that is licensed under the LGPL. The library provides a variety of functionalities intended to be useful for research purposes related to biometric template protection. In particular, it provides the \href{\THIMBLEDOCURL/classthimble_1_1BCHCode.html}{\tt BCHCode} class of which objects represent binary BCH codes which can correct binary vectors, represented by objects from the \href{\THIMBLEDOCURL/classthimble_1_1BinaryVector.html}{\tt BinaryVector} class, to their closest codewords within the BCH code's error-correcting radius. In this appendix we demonstrate how the experiments in Section \ref{sec:experiments} can be reproduced using THIMBLE and its \href{\THIMBLEDOCURL/classthimble_1_1BCHCode.html}{\tt BCHCode} class.

An executable program that uses THIMBLE may be of the following form
\lstset{language=C++}
\begin{lstlisting}
#include <thimble/all.h>

using namespace std;
using namespace thimble;

int main( int argc , char *args[] ) {

   // IMPLEMENT ME

   return 0;
}
\end{lstlisting}
where \verb|// IMPLEMENT ME| should be replaced by C++ code. To generate an $(\TheBlockLength,\TheDimension,\TheMinimalDistance)$=(127,36,31)-BCH code (which is one of the BCH codes considered in Kelkboom \etal{} \cite{bib:KelkboomEtAl2011} being of error-correcting capability $\approx 25\%$), we may run
\begin{lstlisting}
   BCHCode C(127,15);
   
   int n = C.getBlockLength();
   int k = C.getDimension();
   int d = C.getMinimalDistance();
\end{lstlisting}
which creates a BCH code of block length $127$ being capable of correcting (at least) $15$ bit errors and is thus of minimal distance $\TheMinimalDistance=31$. We may choose two feature vectors of Hamming distance 4, say, at random by
\begin{lstlisting}
   BinaryVector w1(n) , w2(n) , e(n);
   int hw = 4;
   
   w1.random(true); // first feature vector
   e.wrandom(hw,true); // error pattern
   w2 = w1 + e; // second feature vector
\end{lstlisting}
where the arguments {\tt true} advise the \href{\THIMBLEDOCURL/classthimble_1_1BinaryVector.html#a9cf9f082e286d67949e93f2962e66e9d}{\tt random()} methods to use a strong random generator (\eg{} {\tt /dev/urandom} on UNIX-based systems); note that the above C++ code can easily be adjusted to simulate the attack in the non-related case by also choosing the second feature set uniformly at random. Two codewords may be randomly drawn from the BCH code by
\begin{lstlisting}
   BinaryVector c1 , c2;
   
   C.random(c1,true);
   C.random(c2,true);
\end{lstlisting}
To generate two random permutation processes, represented by objects from the \href{\THIMBLEDOCURL/classthimble_1_1Permutation.html}{\tt Permutation} class, run
\begin{lstlisting}
   Permutation P1(n) , P2(n);
   
   P1.random(true);
   P2.random(true);
\end{lstlisting}
Next, we can construct two fuzzy commitments.
\begin{lstlisting}
   BinaryVector f1 , f2;
   
   f1 = c1 + P1*w1;
   f2 = c2 + P2*w2;
\end{lstlisting}
Now, assume that we are an adversary who has intercepted the records {\tt (f1,P1)} and {\tt (f2,P2)}; we keep track of correct feature vectors and codewords to be later able to verify the correctness of an output. With the notation of Algorithm \ref{alg:theattack}, we may obtain generator matrices encoded as \href{\THIMBLEDOCURL/classthimble_1_1BinaryMatrix.html}{\tt BinaryMatrix} objects for the codes $\TheCode_1$ and $\TheCode_2$ through
\begin{lstlisting}
   BinaryMatrix G1 , G2;
   
   G1 = inv(P1) * C.getGeneratorMatrix();
   G2 = inv(P2) * C.getGeneratorMatrix();
\end{lstlisting}
and we can compute its concatenation $\tilde{\GenMatrix}$
\begin{lstlisting}
   BinaryMatrix tG = concatCols(G1,G2);
\end{lstlisting}
A check matrix $\tilde{\CheckMatrix}$ for the code generated by $\tilde{\GenMatrix}$ may be computed through
\begin{lstlisting}
   BinaryMatrix tH = transpose(kernel(transpose(tG)));
\end{lstlisting}
Starting from a zero error pattern $\ErrorPattern$ encoded by a \href{\THIMBLEDOCURL/classthimble_1_1BinaryVector.html}{\tt BinaryVector} we may successively iterate through an error pattern sequence of increasing Hamming weight until for the offset $\ReceivedWord=\PermMatrix_1^{-1}\FuzzyCommitment_1-\PermMatrix_2^{-1}\FuzzyCommitment_2$ the equality $\tilde{\CheckMatrix}\cdot(\ReceivedWord-\ErrorPattern)=0$ holds.
\begin{lstlisting}
   BinaryVector r = inv(P1) * f1 - inv(P2) * f2;
   
   // we use the error pattern from above for
   // the guesses of the error pattern
   e.setZero(); 
   
   // Keeps track of whether a working error
   // pattern is found.
   bool success = false;
   
   // Loop as long as the Hamming weight is smaller
   // than 'hw'
   while (e.hammingWeight()<=hw) {
   
      if ( isInKernel(r-e,tH) ) {
          success = true;
          break;
      }
   
      // next element in the sequence of error
      // patterns of increasing Hamming weight
      e.next();
   }
   
   // If no working error pattern has been found,
   // output a message, resembling that the records
   // definitely do not protect feature sets of 
   // Hamming distance smaller than or equals 'hw'
   if ( !success ) {
      cout << "NON-RELATED" << endl;
      return 0;
   }
\end{lstlisting}
After the iteration, we may try to find the feature vectors using the relations described in Section \ref{sec:GeneralizedDecodabilityAttack} and Section \ref{sec:theattack} from the solution of a linear system.
\begin{lstlisting}
   BinaryVector m;

   // Solve the linear system
   solve(m,tG,r-e);

   // Message vectors
   BinaryVector m1(k) , m2(k);
   for ( int i = 0 ; i < k ; i++ ) {
       m1.setAt(i,m.getAt(i));
       m2.setAt(i,m.getAt(i+k));
   }

   // Codeword candidates
   BinaryVector rc1 , rc2;
   rc1 = C.getGeneratorMatrix() * m1;
   rc2 = C.getGeneratorMatrix() * m2;
      
   // Feature vector candidates
   BinaryVector rw1 , rw2;
   rw1 = inv(P1)*(f1 - rc1);
   rw2 = inv(P2)*(f2 - rc2);

   // Check if recovery was successful and
   // output a corresponding message.
   if ( rw1 == w1 && rw2 == w2 ) {
       cout << "REVERTED" << endl;
   } else {
       // If recovery was unsuccessful, then
       // only label the records as related.
       cout << "RELATED" << endl;
   }
\end{lstlisting}
In average, the attack outputs in nearly $50\%$ of the cases {\tt REVERTED} and in the other cases {\tt RELATED}. If the C++ code is adjusted to attack non-related records, then it will with probability very close to $0\%$ output {\tt NON-RELATED}. Note that due to the higher error-correcting capability of $\approx 25\%$ of an $(127,36)$-BCH code the attack performs better as for the $(127,50)$-BCH code considered in Table \ref{tab:experiments} which can correct up to $\approx 10\%$ errors but for which the attack fails in $\approx 4.18\%$. We stress that the C++ code example given in this appendix can be easily adjusted to reproduce our experimental results from Section \ref{sec:experiments}.

\end{document}